\documentclass[useAMS,usenatbib]{mn2e}
\usepackage{epsfig}

\begin{document}

\title{Dynamical Friction in a magnetized gas}

\author[ M. Shadmehri \& F. Khajenabi]{Mohsen Shadmehri\thanks{Email:m.shadmehri@gu.ac.ir} and Fazeleh Khajenabi\thanks{E-mail:
f.khajenabi@gu.ac.ir;}\\
School of Physics, Faculty of Science, Golestan University, Gorgan, Iran}

\maketitle

\date{Received ______________ / Accepted _________________ }

\begin{abstract}
When a gravitating point mass moves subsonically  through a magnetized and isothermal medium, the dynamical structure of the flow is studied far from the mass using a perturbation analysis. Analytical solutions for the first-order density and the velocity perturbations  are   presented. Validity of our solutions is restricted to the cases where   the Alfven velocity  in the ambient medium is less than the accretor's velocity. The density field is less dense because of the magnetic effects according to the solutions and the dynamical friction force becomes lower as the strength of the magnetic field increases.

\end{abstract}

\section{Introduction}

In many astronomical systems when an object (e.g., a star ) travels through a collisionless or a collisional system, it is decelerated in the direction of its motion because of the gravitational interaction of the perturber (i.e, the object) and the wakes it excites in the ambient medium. Theoretical efforts to study this problem have been started after a pioneering work by \citet{chandra43}. He considered a collisionless background and the drag force that the perturber experiences was found analytically.  Since then it has been applied to many astronomical systems, as in  migration of protoplanets in protoplanetary accretion discs \citep*[e.g.,][]{muto2011,livio2009}, motion of supermassive black holes within gas-rich galaxies \citep*[e.g.,][]{sij,ruder71}, orbital evolution of embedded binary stars \citep*[e.g.,][]{stah2010}, motion of compact stars around supermassive black holes  \citep{narayan2000},  and the heating of the galaxy clusters \citep*{kim2005,kim2004} and even studies at the cosmological scales \citep{tittley2001}.

The standard approach based on the linearized hydrodynamics equations for describing the excited small amplitude perturbations and calculating their gravitational interaction with the  object moving in a straight-line trajectory \citep[e.g.,][]{dokuch,salp80,ostriker99}. Numerical simulations confirmed results of the linear analysis for the drag force
\citep[e.g.,][]{sanchez99,kimkim2009}. The standard approach has  been extended to the cases with circular-orbit perturber \citep{kim2010}, double perturbers  \citep{kimkim2008} or accelerated motion in a straight-line \citep{namo2010}. In calculating the drag force, the role of the relativistic effects has also been studied by \citet{barausse}. A modified Newtonian approach is used by \citet{sanchez2009} to analysis the drag force. \\

Recently, \citet{LS} (hereafter LS) argued that the physical size of the perturber is much smaller than the accretion radius in the astronomical systems such as  supermassive black holes within galaxies or giant planets inside protoplanetary discs. Under these circumstances, one can not neglect the mass accretion and the amplitude of the wakes is not small as authors have assumed in their linear approach for calculating the drag force. Thus, to a large extend it is the mass accretion which provides the linear momentum transfer from the ambient medium to the object. LS showed analytically that the steady-sate friction force is equal to the accretion rate onto the object  multiply by its velocity. If the accretion rate is prescribed by the Bondi-Hoyle rate \citep{BondiHoyle}, the drag force does not rise monotonically but instead peaks around Mach number $0.68$ and then begin to decline (LS). It is an interesting result that may find application in a number of astronomical systems. However, there are many other important physical factors that may modify the drag force, and, we think, the approach of LS has this possibility to be generalized  to include  those physical ingredients such as a magnetized and/or isentropic ambient medium. Just recently, \citet{khajenabi2012} extended the approach of LS to the case of an isentropic medium and found a set of analytical solutions for  the structure of the flow far from object. They proved the proportionality between the friction force and the accretion rate also holds in the isentropic case.

The purpose of this paper is to calculate the dynamical friction force when the perturber  moves with constant velocity through a {\it magnetized } medium. In most of the astronomical systems, the ambient medium is actually magnetized, and one may expect the excited wakes are modified because of the magnetic effects. In performing this analysis, we also assume the physical size of the object is much smaller than the accretion radius and the approach of LS is followed, but including the magnetic fields. Basic equations are presented in the next section. Introducing perturbation expansions in section 3, we will find analytical solutions for the first-order perturbed variables in section 4. However, second-order solutions and the dynamical friction force are obtained numerically in section 5. We conclude by some physical implications of the results in section 6.

\section{basic equations}
\begin{figure}
\vspace{-36pt}
\epsfig{figure=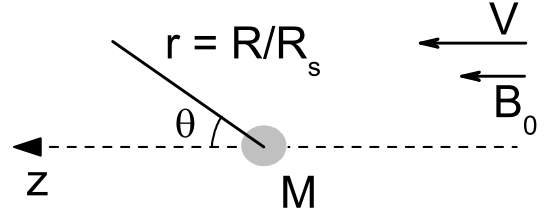,angle=0,scale=0.65}
\caption{This figure shows the assumed magnetic field and the flow velocity. The object is fixed at the origin and the magnetic field and the velocity of the gas are both  uniform at the large distances from the object.}
\label{fig:fnew}
\end{figure}

The basic equation of our problem in the steady state are written as
\begin{equation}
\nabla \cdot (\rho {\bf U}) =0,
\end{equation}
\begin{equation}
\rho ({\bf U} \cdot \nabla) {\bf U} = -\nabla P - \rho \frac{GM}{R^2} + \frac{1}{4\pi} {\bf J} \times {\bf B},
\end{equation}
\begin{equation}
\nabla \times ({\bf U} \times {\bf B})=0,
\end{equation}
\begin{equation}
\nabla \cdot {\bf B} =0,
\end{equation}
where $\rho$, ${\bf U}$, $P$ are the density, the velocity and the pressure, respectively. Here, $R$ is the radial distance in the spherical coordinates $(R,\theta,\varphi)$ whose origin is anchored on the gravitating mass $M$ (see Figure \ref{fig:fnew}). The current density is ${\bf J}=\nabla \times {\bf B}$. Moreover, we also assume the gas is isothermal, i.e. $P=\rho c_{\rm s}^{2}$ where $c_{\rm s}$ is the sound speed. We ignore the complications from density gradients. Very far from the mass which travels in a straight line trajectory with velocity $V$ not only the density, but the magnetic fields are both assumed to be spatially uniform, i.e. $\lim_{R\rightarrow \infty} \rho \equiv \rho_0$ and $\lim_{R \rightarrow \infty} {\bf B} \equiv B_{0} {\bf e}_{\rm z}$ where ${\bf e}_{\rm z}$ is the unit vector along the $z$ direction (Figure \ref{fig:fnew}).

Introducing a set of non-dimensional variables as $r=R/R_{\rm s}$, ${\bf u} = {\bf U} / c_{\rm s}$, $\varrho = \rho/\rho_{0}$ and ${\bf b}={\bf B}/B_{0}$, we can write our main equations in the non-dimensional forms. Here, the sonic radius is defined as $R_{\rm s} \equiv GM/c_{\rm s}^2$. Therefore, we have
\begin{equation}\label{eq:con}
\nabla \cdot (\varrho {\bf u})=0,
\end{equation}
\begin{equation}\label{eq:momentum}
\varrho ({\bf u} \cdot \nabla ) {\bf u} = - \nabla \varrho - \frac{\varrho}{ r^2} + \xi^{2} (\nabla \times {\bf b}) \times {\bf b},
\end{equation}
\begin{equation}\label{eq:induction}
\nabla \times ({\bf u} \times {\bf b}) =0,
\end{equation}
\begin{equation}\label{eq:divergence}
\nabla \cdot {\bf b} =0,
\end{equation}
where the non-dimensional parameter $\xi$ is defined as the ratio of the Alfven velocity $V_{\rm A} = B_{0}/\sqrt{4\pi\rho_{0}}$ and the sound speed, i.e. $\xi = V_{\rm A} / c_{\rm s}$. Obviously, in equations (\ref{eq:con})-(\ref{eq:divergence}) the radial derivatives are calculated with the respect to the non-dimensional radial distance $r$. We also define $\beta = V/c_{\rm s}$. Thus, properties of the flow at the distances far from the gravitating object $M$ are ${\bf u} = \beta {\rm e}_{\rm z}$, $\varrho = 1$ and ${\bf b} = {\bf e}_{\rm z}$ when $r\gg 1$.

Note that in our subsequent calculations, it is assumed that  the system is axisymmetric ($\partial / \partial\varphi =0$). If we introduce the non-dimensional stream function $\psi (r,\theta)$ as
\begin{equation}
\varrho {\bf u} = \nabla \times (\frac{\psi}{ r \sin\theta} {\bf e}_{\varphi}),
\end{equation}
then the non-dimensional continuity equation (\ref{eq:con}) is automatically satisfied. Thus, the components of the velocity can be written in terms of the
stream function as
\begin{equation}
u_{\rm r} = \frac{1}{\varrho r^2 \sin\theta} \frac{\partial\psi}{\partial\theta},
\end{equation}
\begin{equation}
u_{\theta} = \frac{-1}{\varrho r \sin\theta} \frac{\partial\psi}{\partial r}.
\end{equation}
We can also define the non-dimensional positive-definite magnetic flux function $\Phi$ as
\begin{equation}
{\bf b} = \nabla \times (\frac{\Phi}{r \sin\theta} {\bf e}_{\varphi}).
\end{equation}
So, equation (\ref{eq:divergence}) is satisfied and the components of the magnetic field become
\begin{equation}
b_{\rm r} = \frac{1}{ r^2 \sin\theta} \frac{\partial\Phi}{\partial\theta},
\end{equation}
\begin{equation}
b_{\theta} = \frac{-1}{ r \sin\theta} \frac{\partial\Phi}{\partial r}.
\end{equation}

Having the above equations for the components of the velocity in terms of $\psi$ and the magnetic field in terms of  $\Phi$, we are left only with the momentum equation (\ref{eq:momentum}) and the induction equation (\ref{eq:induction}). Thus, we have
\begin{displaymath}
\varrho \left ( u_{\rm r} \frac{\partial u_{\rm r}}{\partial r} + \frac{u_{\theta}}{r} \frac{\partial u_{\rm r}}{\partial\theta} - \frac{u_{\theta}^2}{r} \right ) = - \frac{\partial \varrho}{\partial r} - \frac{\varrho}{r^2} - \xi^2 \frac{b_{\theta}}{ r} \times
\end{displaymath}
\begin{equation}\label{eq:momR}
\left [ \frac{\partial}{\partial r} (r b_{\theta}) - \frac{\partial b_{\rm r}}{\partial \theta } \right ],
\end{equation}
\begin{displaymath}
\varrho \left ( u_{\rm r} \frac{\partial u_{\theta}}{\partial r} + \frac{u_{\theta}}{r} \frac{\partial u_{\theta}}{\partial\theta} + \frac{u_{\rm r} u_{\theta}}{r} \right ) = -\frac{1}{r} \frac{\partial\rho}{\partial\theta} + \xi^2 \frac{b_{\rm r}}{r} \times
\end{displaymath}
\begin{equation}\label{eq:momtet}
\left [ \frac{\partial}{\partial r} (r b_{\theta}) - \frac{\partial b_{\rm r}}{\partial \theta } \right ],
\end{equation}
\begin{equation}\label{eq:sinduction}
u_{\rm r} b_{\theta} - u_{\theta} b_{\rm r} =0.
\end{equation}
Since we have neglected the dissipative terms in the induction equation and the system is steady state, we could say based on the equations (\ref{eq:con}) and (\ref{eq:divergence}) that the magnetic field and the velocity vectors are parallel. The simplified induction equation (\ref{eq:sinduction}) also shows this behavior.

\begin{figure}
\epsfig{figure=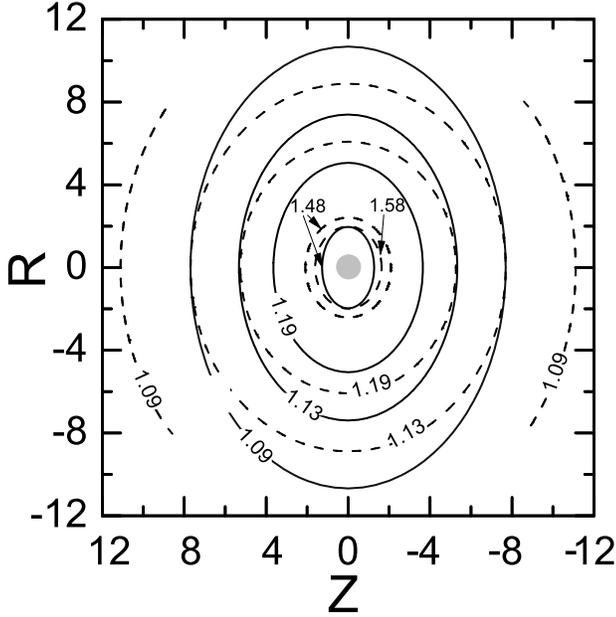,angle=0,scale=0.5}
\caption{Isodensity curves corresponding to the first-order density perturbation for $\beta =0.5$ and $\xi =0.4$ ({\it solid}) and $\xi=0$ ({\it dashed}). Each contour is labeled by its level. The gray inner circle denotes the sonic radius. Obviously, the solutions are more accurate at the distances far from the inner sonic circle.}
\label{fig:first-iso}
\end{figure}

\begin{figure}
\epsfig{figure=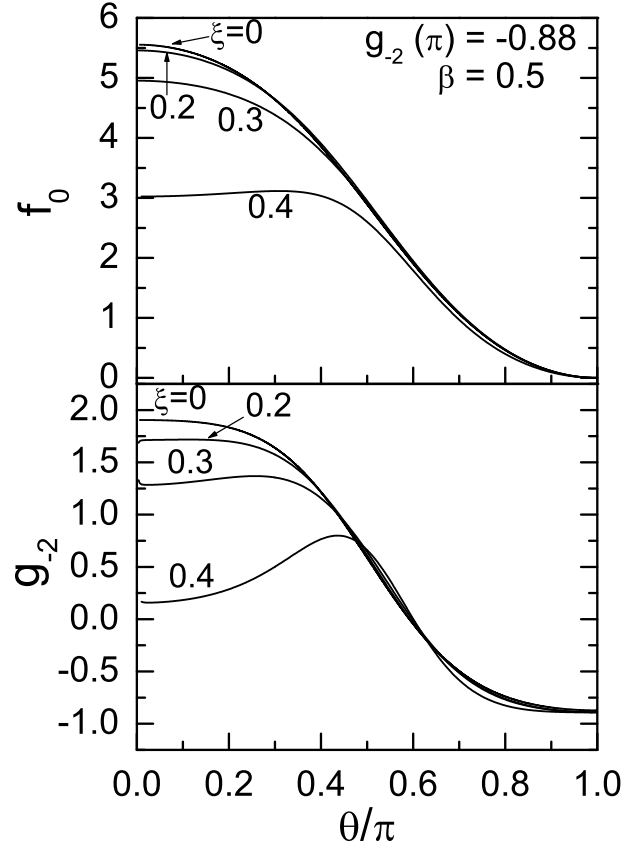,angle=0,scale=0.5}
\caption{Profiles of $f_0 (\theta)$ and $g_{-2} (\theta) $ for different values of $\xi$ when the initial value of $g_{-2}(\pi )$ is -0.88. Each curve is labeled by the corresponding value of $\xi$.}
\label{fig:g88}
\end{figure}

\begin{figure}
\epsfig{figure=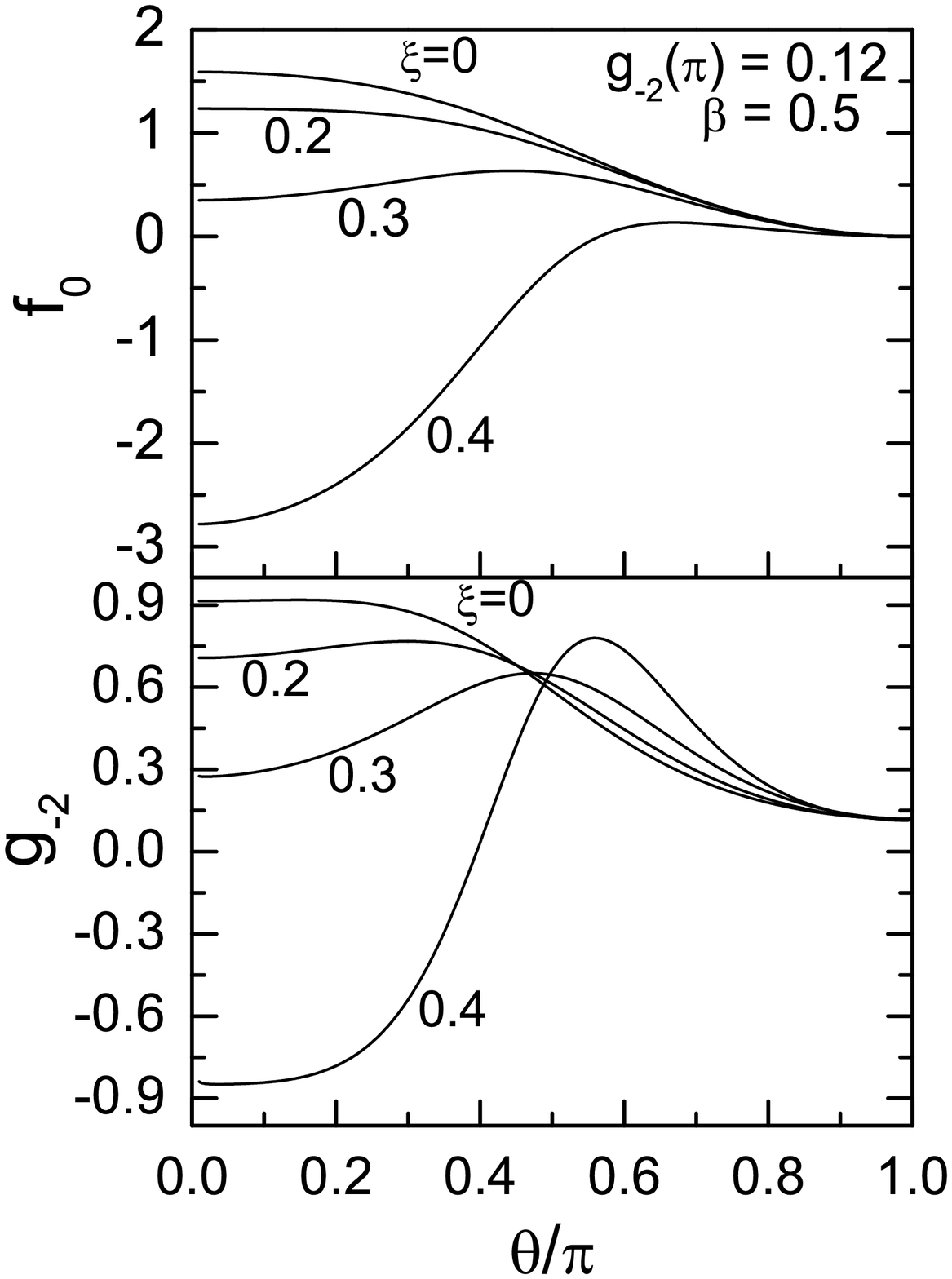,angle=0,scale=0.5}
\caption{The same as Figure \ref{fig:g88}, but for $g_{-2}(\pi )=0.12$.}
\label{fig:g12}
\end{figure}

\begin{figure}
\epsfig{figure=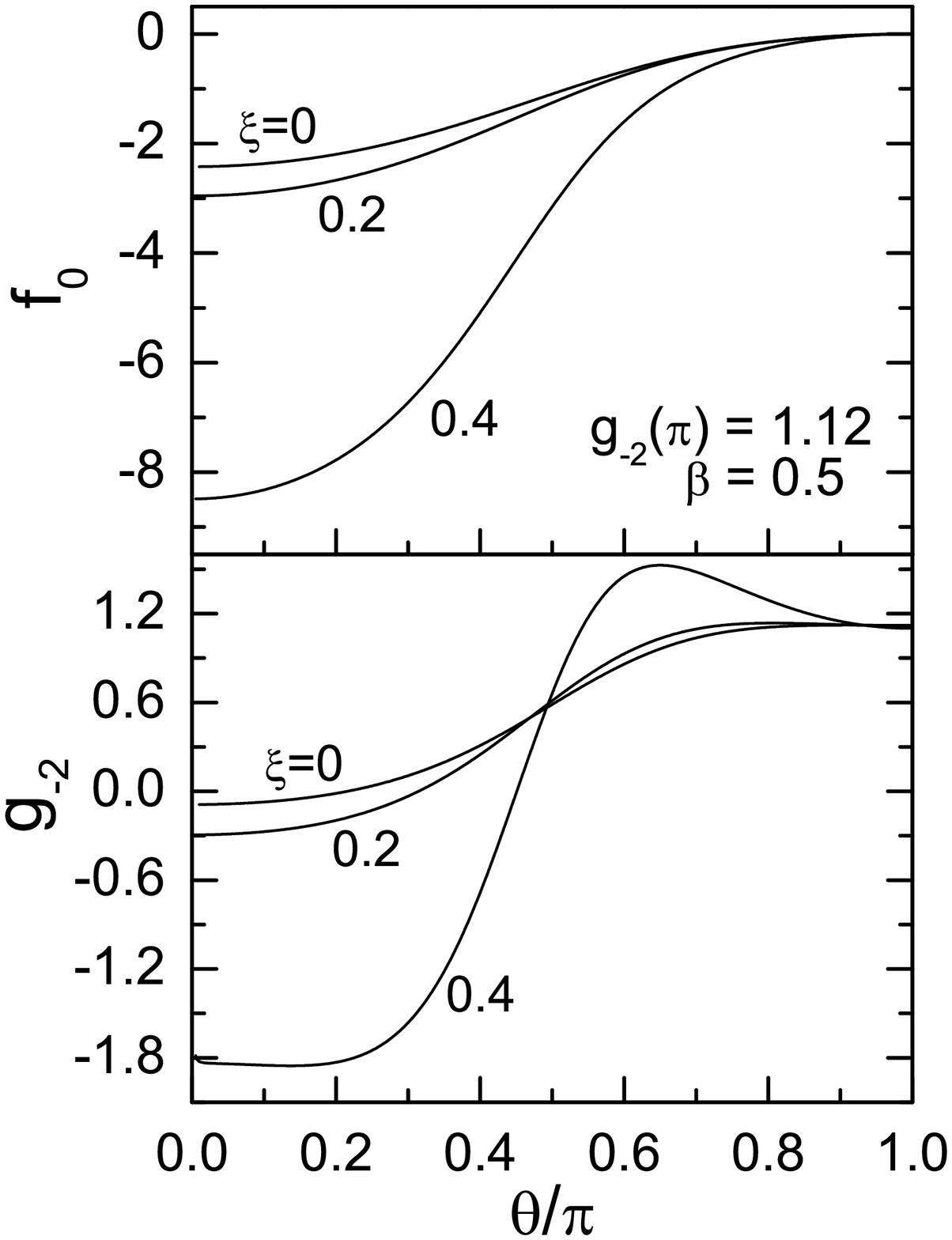,angle=0,scale=0.5}
\caption{The same as Figure \ref{fig:g88}, but for $g_{-2}(\pi )=1.12$.}
\label{fig:g112}
\end{figure}

\section{perturbation expansions}
We expand the density and the stream function similar to LS as
\begin{equation}\label{eq:density}
\varrho (r,\theta) = 1 + g_{-1} (\theta) r^{-1} + g_{-2} (\theta) r^{-2} + g_{-3} (\theta) r^{-3} + \cdots ,
\end{equation}
\begin{equation}\label{eq:psi}
\psi (r,\theta) = f_{2}(\theta) r^2 + f_1 (\theta) r + f_0 (\theta) + f_{-1}(\theta) r^{-1} +  \cdots ,
\end{equation}
and in our magnetized model, the magnetic flux function is expanded as
\begin{equation}\label{eq:Phi}
\Phi (r,\theta)= h_{2}(\theta) r^2 + h_1 (\theta) r + h_0 (\theta) + h_{-1} (\theta) r^{-1} + \cdots .
\end{equation}

Since the velocity tends to be constant far from the object, one can simply obtain $f_{2}(\theta )=(1/2)\beta \sin^2\theta$. Also, magnetic field lines are uniform far from the object and then, $h_{2}(\theta )=(1/2) \sin^2\theta$. We can set $\psi (r, \pi )=0$, and so $f_{i}(\pi )=0$ for $i=1,0,-1,-2$ etc., and regularity of the components of the velocity on the axis defined by $\theta =0$ and $\theta =\pi$ implies $f_{i}'(\pi ) = f_{i}'(0)=0$, for $i=1,0,-1,-2,$ etc., and $f_{i}(0)=0$ for $i=1,-1,-2,$ etc. (see LS for details of their arguments). Boundary conditions on the geometry of the magnetic field lines also imply $h_{i}(\pi )=h_{i}'(\pi ) = h_{i}'(0)=0$, for $i=1,0,-1,-2,$ etc., and $h_{i}(0)=0$ for $i=1,-1,-2,$ etc. Substituting the above expansions into the main equations, a set of ordinary differential equations are obtained, which are solvable analytically and/or numerically subject to the mentioned  boundary conditions.

Also, the mass accretion rate is written as
\begin{equation}
\dot{{\cal M}}=-2\pi \int_{0}^{\pi} \rho U_{\rm R} R^2 \sin\theta d\theta .
\end{equation}
If we nondimensionalize the accretion rate $\dot{{\cal M}}$ by $2\pi \rho_0 c_{\rm s} R_{\rm s}^2$ and using expansion (\ref{eq:psi}), the nondimensonal accretion rate $\dot{M}$ becomes
(see LS for the details)
\begin{equation}
{\dot M} = f_{0}(0).
\end{equation}
\section{First order equations}
We can now substitute the expansions (\ref{eq:density}), (\ref{eq:psi}) and (\ref{eq:Phi}) into the momentum equations (\ref{eq:momR}) and (\ref{eq:momtet}) and the induction equation (\ref{eq:sinduction}). Then, we match the coefficients of
each power of $r$. Like the non-magnetized case (LS), we find the highest power of $r$ is $r^{-1}$, but all the corresponding coefficients on both sides of the equations vanish identically.  So, we consider the next power of $r$ (i.e. $r^{-2}$) and the first-order equations are obtained. From the radial component of the momentum equation, we have
\begin{displaymath}
-\beta f_{1}'' -\beta f_{1} + \beta^{2} \sin\theta \cos\theta g_{-1}' + (\beta^{2}\cos^2\theta -1) g_{-1} +
\end{displaymath}
\begin{equation}\label{eq:first-r}
1 + \xi^2 ( h_{1}'' - \cot\theta h_{1}' ) =0,
\end{equation}
and the $\theta-$component of the momentum equation gives
\begin{displaymath}
(1-\beta^2 \sin^{2}\theta) g_{-1}' - \beta^{2} \sin\theta \cos\theta g_{-1} +
\end{displaymath}
\begin{equation}\label{eq:first-tet}
\xi^2 \cot\theta ( h_{1}'' - \cot\theta h_{1}' ) =0,
\end{equation}
and finally the induction equation (\ref{eq:sinduction}) yields
\begin{equation}\label{eq:first-induction}
 \cos\theta f_1 - \beta\cos\theta h_1 + \beta \sin\theta h_{1}' - \sin\theta f_{1}'=0
\end{equation}

For the non-magnetized gas, the above first-order equations reduce to the equations of LS if we set $\xi =0$. We can solve these equations for $f_{1}$, $g_{-1}$ and $h_1$, and then properties of the flow to the first order are determined. One can integrate equation (\ref{eq:first-induction}) simply by re-arranging its terms as
\begin{equation}
 \cos\theta (f_1 - \beta h_1 ) - \sin\theta \frac{d}{d\theta} (f_1 - \beta h_1 )=0.
\end{equation}
Introducing $W=f_1 - \beta h_1$, this equation becomes $dW/d\theta = W \cot\theta$ and its solution is $W=W_0 \sin\theta $ where $W_0$ is an arbitrary constant to be determined from the boundary conditions. Thus, equation $f_1 - \beta h_1= W_0 \sin\theta$ is valid for the whole range of $\theta$ including $\theta = \pi$. So, we have $f_1'(\pi) - \beta h_1'(\pi) = W_0 \cos\pi = -W_0$. Considering our imposed boundary conditions $f_1'(\pi)=h_1'(\pi)=0$, we obtain  $W_0 = 0$ and
\begin{equation}\label{eq:induc-f}
f_1 = \beta h_{1}.
\end{equation}

We can now substitute equation (\ref{eq:induc-f}) into equations (\ref{eq:first-r}) and (\ref{eq:first-tet}) and the following equations for $f_1$ and $g_{-1}$ are obtained
\begin{displaymath}
(\frac{\xi^2 - \beta^2 }{\beta}) f_{1}'' - \frac{\xi^2}{\beta} \cot\theta f_{1}' - \beta f_{1} + \beta^2 \sin\theta\cos\theta g_{-1}' +
\end{displaymath}
\begin{equation}\label{eq:first-main1}
(\beta^2 \cos^{2}\theta -1) g_{-1} + 1 =0,
\end{equation}
\begin{displaymath}
(1-\beta^2 \sin^2 \theta) g_{-1}' - \beta^2 \sin\theta\cos\theta g_{-1} + \frac{\xi^2}{\beta} \cot\theta f_{1}'' -
\end{displaymath}
\begin{equation}\label{eq:first-main2}
\frac{\xi^2}{\beta} \cot^2\theta f_{1}' =0.
\end{equation}

Thus, equations (\ref{eq:first-main1}) and (\ref{eq:first-main2}) are our main equations to be solved. Interestingly, these equations are integrable, though  the mathematical manipulation is cumbersome. Here, we summarize the solutions as
\begin{equation}\label{eq:f1}
f_{1}(\theta ) = \frac{1}{\beta} - \frac{1}{\beta} \frac{1}{\sqrt{\Gamma} } \sqrt{\Gamma - \beta^2 \sin^2 \theta },
\end{equation}
\begin{equation}\label{eq:g-1}
g_{-1}(\theta )= \frac{1-\xi^2 /\beta^2}{\sqrt{\Gamma}} \frac{1}{\sqrt{\Gamma - \beta^2 \sin^2 \theta }},
\end{equation}
where $\Gamma = 1+ \xi^2 - \xi^2 / \beta^2 $. For the non-magnetized case (i.e. $\xi =0$), we have $\Gamma = 1$  and the above solutions reduce to equations (25) and (26) of LS. Note that the above first order solutions are physically acceptable if $\Gamma - \beta^2 \sin^2 \theta  > 0$.  If for $\theta =\pi /2$ (where the second term has a maximum value) the left hand side becomes positive, then for other values of $\theta$ the left-hand side  would be positive (as we want). Having $\Gamma = 1 + \xi^2 - \xi^2 / \beta^2 $, we can write $1 + \xi^2 - \xi^2 / \beta^2 - \beta^2 > 0$ which implies $\beta > \xi$. Thus, our solutions are applicable to the case where the Alfven speed is less than the sound speed.

\begin{figure}
\epsfig{figure=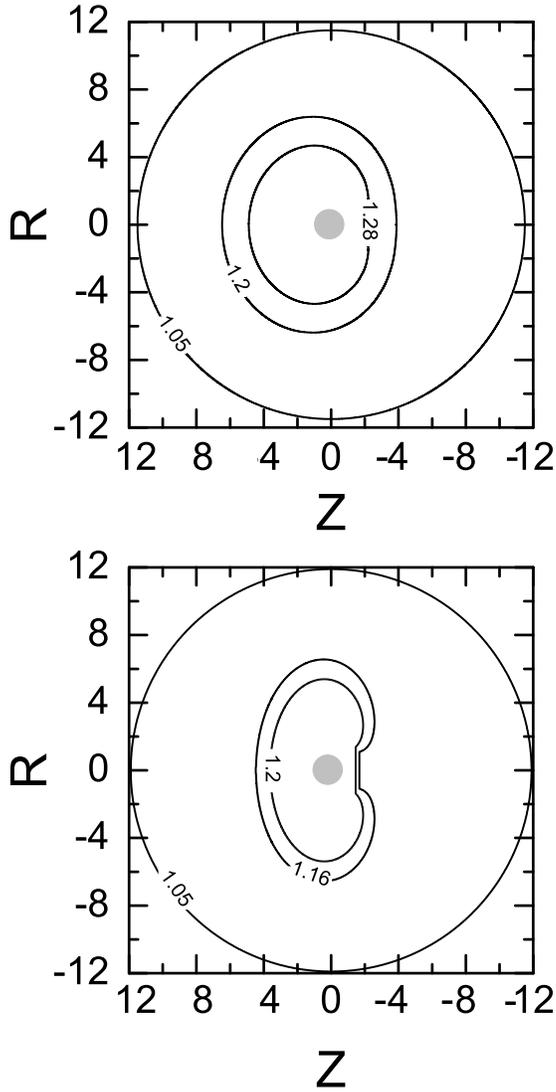,angle=0,scale=0.5}
\caption{Density contours for the non-magnetized ({\it top plot}) and the magnetized flow ({\it bottom plot}) with $\beta =0.5$, $\xi =0.4$ and $g_{-2}(\pi ) =-0.88$.}
\label{fig:density-second}
\end{figure}

\begin{figure}
\epsfig{figure=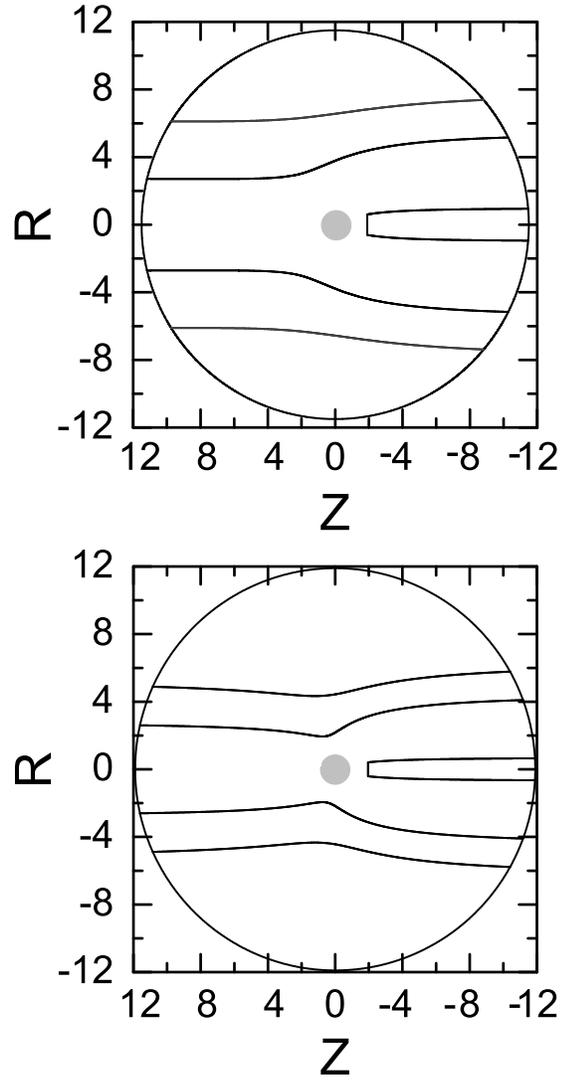,angle=0,scale=0.5}
\caption{Streamlines for the non-magnetized ({\it top plot}) and the magnetized flow ({\it bottom plot}) with $\beta =0.5$, $\xi =0.4$ and $g_{-2}(\pi ) =-0.88$.}
\label{fig:velocity-second}
\end{figure}

Figure \ref{fig:first-iso} shows isodensity contours for the first-order density perturbation, i.e. equation (\ref{eq:g-1}). Solid contours are for a magnetized ambient medium when $\xi =0.4$. Nonmagnetized contours are also shown by the dashed curves. In both cases, the Mach number is fixed $\beta =0.5$.  A circle with unit radius at the center denotes the sonic radius. Obviously, our solutions are valid beyond the sonic radius and all the contours are drawn for the radii larger than 2. Each contour is labeled by the corresponding level. For the non-magnetized solutions, the density varies from $1.09$ to $1.58$ at the inner parts and the magnetic effects change this interval from $1.09$ to $1.48$. In other words, the density field in the presence of the magnetic field is less dense in comparison to the non-magnetic solutions. When magnetic fields are not considered, the perturbations in the ambient medium grow easier and so the wakes have larger amplitudes comparing to the magnetic case. It is because of the magnetic pressure which acts as an extra pressure against further growth of the perturbations.

\section{second-order equations}
Now, we can equate coefficients of the next power of $r$ in the components of the momentum equation and the induction equation. From the induction equation, we obtain
\begin{equation}
f_0 = \beta h_0.
\end{equation}

Considering the above relation between $f_0$ and $h_0$, we obtain the following equation from the radial component of the momentum equation
\begin{displaymath}
(\frac{\xi^2}{\beta}-\beta) f_{0}'' - (\beta + \frac{\xi^2}{\beta}) \cot\theta f_{0}' + \beta^{2} \sin\theta \cos\theta g_{-2}' +
\end{displaymath}
\begin{equation}\label{eq:second-r}
(2\beta^{2} \cos^{2}\theta - 2) g_{-2} = {\cal A}_{1} + {\cal A}_{2} + {\cal A}_{3} + {\cal A}_{4},
\end{equation}
where all the terms on the righthand side depend on the first order variables as
\begin{equation}
{\cal A}_{1}= \frac{f_{1}^2}{\sin^2 \theta} - \frac{f_{1}f_{1}' \cos\theta }{\sin^{3}\theta} + \frac{(f_{1}')^2}{\sin^2 \theta} + \frac{f_{1} f_{1}''}{\sin^2 \theta},
\end{equation}
\begin{displaymath}
{\cal A}_{2} = \beta f_1 g_{-1} - 2\beta f_{1}' g_{-1} \cot\theta - \beta f_{1} g_{-1}' \cot\theta -
\end{displaymath}
\begin{equation}
\beta f_{1}' g_{-1}' + \beta f_{1}'' g_{-1},
\end{equation}
\begin{equation}
{\cal A}_{3} = 2 (g_{-1})^2 - 3 g_{-1}.
\end{equation}
\begin{displaymath}
{\cal A}_{4} = (\frac{\xi^2}{\beta})(-2 f_{1}'' g_{-1} + 2\cot\theta f_{1}' g_{-1} + \frac{f_{1}' f_{1} \cot\theta }{\beta \sin^{2}\theta} -
\end{displaymath}
\begin{equation}
\frac{f_{1}'' f_{1}}{\beta\sin^{2}\theta})
\end{equation}
Also, the $\theta-$component of equation of motion gives
\begin{displaymath}
(\frac{\xi^2}{\beta})  \cot\theta f_{0}''-(\beta + \frac{\xi^2}{\beta} \cot^{2}\theta ) f_{0}' + (1-\beta^2 \sin^2\theta) g_{-2}' -
\end{displaymath}
\begin{equation}\label{eq:second-theta}
 2\beta^2 \sin\theta \cos\theta g_{-2} = {\cal B}_{1}+{\cal B}_{2}+{\cal B}_{3}+{\cal B}_{4},
\end{equation}
where
\begin{equation}
{\cal B}_{1}= f_1^{2} \frac{\cot\theta}{\sin^2\theta} - \frac{f_{1} f_{1}'}{\sin^2\theta},
\end{equation}
\begin{equation}
{\cal B}_{2} = \beta f_{1} g_{-1} \cot\theta + \beta f_{1}' g_{-1} + 2\beta f_{1} g_{-1}',
\end{equation}
\begin{equation}
{\cal B}_{3}=-2g_{-1} g_{-1}' ,
\end{equation}
\begin{displaymath}
{\cal B}_{4} = \frac{\xi^2}{\beta^2} ( f_{1}'^{2} \frac{\cot\theta}{\sin^{2}\theta } - \frac{f_{1}' f_{1}''}{\sin^{2}\theta} - 2\beta f_{1}'' g_{-1} \cot\theta  +
\end{displaymath}
\begin{equation}
 2\beta f_{1}'g_{-1} \cot^{2}\theta ).
\end{equation}

Both righthand sides of equations (\ref{eq:second-r}) and (\ref{eq:second-theta}) are determined analytically because we have already found $f_1$ and $g_{-1}$, i.e. equations (\ref{eq:f1}) and (\ref{eq:g-1}). Although equation (\ref{eq:second-theta}) is integrable as we will show, it is very unlikely to solve this equation and equation (\ref{eq:second-r}) for $f_{0}(\theta)$ and $g_{-2}(\theta)$ analytically. So, one should solve these equations numerically subject to the appropriate boundary conditions. Note that if we set $\xi =0$, these equations reduce to equations (27) and (31) of LS. Introducing ${\cal D}= \Gamma - \beta^2 \sin^2 \theta $, we can re-write the  second-order equation (\ref{eq:second-r})  as

\begin{displaymath}
(\frac{\xi^2}{\beta}-\beta) f_{0}'' - (\beta + \frac{\xi^2}{\beta}) \cot\theta f_{0}' + \beta^{2} \sin\theta \cos\theta g_{-2}' +
\end{displaymath}
\begin{displaymath}
(2\beta^{2} \cos^{2}\theta - 2) g_{-2} = \frac{1}{-\Gamma (\Gamma - {\cal D}) {\cal D}^{2} (\beta^2 -1)^2} \times
\end{displaymath}
\begin{equation}\label{eq:second-r-2}
({\cal R}_{1} {\cal D}^{5/2} + {\cal R}_{2} {\cal D}^2 + {\cal R}_{3} {\cal D}^{3/2} + {\cal R}_{4} {\cal D} + {\cal R}_{5} ),
\end{equation}
where
\begin{equation}
{\cal R}_{1} =  \sqrt{\Gamma} (1- \beta^2) (\beta^2 + 2 - 3 \Gamma ),
\end{equation}
\begin{equation}
{\cal R}_{2} = 2 \Gamma^3 - 4 \beta^2 \Gamma^2 + (5 \beta^2 -3) \Gamma + \beta^2 - \beta^4 ,
\end{equation}
\begin{equation}
{\cal R}_{3}= 3 \Gamma^{3/2} (\beta^2 - 1) (\beta^2 - \Gamma ),
\end{equation}
\begin{equation}
{\cal R}_{4} = \Gamma (\Gamma - \beta^2 ) [ - 4 \Gamma^2 + 2(1+2\beta^2 )\Gamma + 1- 3\beta^2 ],
\end{equation}
\begin{equation}
{\cal R}_{5} = 2 \Gamma^2 (\Gamma -1) (\beta^2 - \Gamma )^2 .
\end{equation}
Also, equation (\ref{eq:second-theta}) is written as
\begin{displaymath}
(\frac{\xi^2}{\beta})  \cot\theta f_{0}''-(\beta + \frac{\xi^2}{\beta} \cot^{2}\theta ) f_{0}' + (1-\beta^2 \sin^2\theta) g_{-2}' -
\end{displaymath}
\begin{displaymath}
 2\beta^2 \sin\theta \cos\theta g_{-2} = \frac{\beta^2 \sin\theta \cos\theta}{-\Gamma (\Gamma - {\cal D})^2 {\cal D}^2 (\beta^2 -1)^2} \times
\end{displaymath}
\begin{equation}\label{eq:second-theta-2}
({\cal R}_6 {\cal D}^{5/2} + {\cal R}_7 {\cal D}^2 + {\cal R}_8 {\cal D}^{3/2} + {\cal R}_9 {\cal D} +  {\cal R}_{10}{\cal D}^{1/2} + {\cal R}_{11} ),
\end{equation}
where
\begin{equation}
{\cal R}_{6}= \sqrt{\Gamma} (\Gamma -1) (\beta^2 - 1),
\end{equation}
\begin{equation}
{\cal R}_{7} = 2 \Gamma^3 - 4\beta^2 \Gamma^2 + (5\beta^2 -3) \Gamma + \beta^2 - \beta^4 ,
\end{equation}
\begin{equation}
{\cal R}_{8} = \Gamma^{3/2} (4 \beta^2 - 1 - 3\Gamma ) (\beta^2 -1),
\end{equation}
\begin{equation}
{\cal R}_{9}= \Gamma (\Gamma - \beta^2 ) [-4\Gamma^2 + (5\beta^2 -1) \Gamma + 2 - 2\beta^2],
\end{equation}
\begin{equation}
{\cal R}_{10}= 2 \Gamma^{5/2} (\Gamma - \beta^2  ) (\beta^2 -1 ),
\end{equation}
\begin{equation}
{\cal R}_{11}= \Gamma^2 (\beta^2 - \Gamma ) [-2 \Gamma^2 + (3\beta^2 -1)\Gamma + 1-\beta^2].
\end{equation}

%

The lefthand side of the second-order equation (\ref{eq:second-theta-2}) is perfect derivative. By integrating both side of this equation from $\theta =\pi$ to $0$, we find
\begin{displaymath}
\left [ \frac{\xi^2}{\beta } \cot\theta f'_{0} + (\frac{\xi^2}{\beta} -\beta ) f_0 + {\cal D} g_{-2} \right ]_{\theta = \pi} =
\end{displaymath}
\begin{equation}
\left [ \frac{\xi^2}{\beta } \cot\theta f'_{0} + (\frac{\xi^2}{\beta} -\beta ) f_0 + {\cal D} g_{-2} \right ]_{\theta = 0} .
\end{equation}
Since $f'_{0}(\pi )=f'_{0}(0)=0$, the above equation gives
\begin{equation}
f_{0}(0) = \frac{\beta}{\beta^2 - \xi^2} \left [ g_{-2}(0) - g_{-2}(\pi ) \right ].
\end{equation}
If we set $\xi =0$, the above equation reduces to equation (46) of LS. Here, like to the non-magnetized case, the accretion rate is proportional to the difference, upstream and downstream, of the second-order density perturbation. However, the proportionality constant not only depends on the velocity of the accretor, but also it depends on the magnetic field strength.

We can solve the equations for the same three values of $g_{-2}(\pi)$ as were used by LS, just in order to make easier comparison to the non-magnetized case. Note that our main equations (\ref{eq:second-r-2}) and (\ref{eq:second-theta-2}) are singular at both endpoints $\theta=0$ and $\theta =\pi$. Figures \ref{fig:g88}, \ref{fig:g12} and \ref{fig:g112} show behaviors of $f_{0}(\theta )$ and $g_{-2}(\theta )$ for three fiducial values of $g_{-2} (\pi ) = -0.88, 0.12, 1.12$. Just to make an easier comparison to the non-magnetized profiles of LS these boundary values for  $g_{-2}(\pi )$ have been selected. In each figure, the starting value of $g_{-2}(\pi )$ is fixed, and we play with the ratio $\xi$ as a free input parameter. Profiles of the second-order coefficients strongly depend on the strength of the magnetic field as shown in these figures. The non-magnetic curves passes through the same point at $\theta = \pi /2$. But magnetic effects change this trend, in particular, when $\xi$ becomes closer to $\beta$. However, the coefficient $g_{-2}(\theta)$ is flat at $\theta =0$ and $\pi$ which means the second-order density perturbation is uniform at both the downstream and upstream of the flow irrespective of the existence of the magnetic fields. It is understandable based on our imposed boundary conditions. Density contours and the velocity streamlines are also shown in Figures \ref{fig:density-second} and \ref{fig:velocity-second} for the magnetized and the non-magnetized case corresponding to the parameters $\beta=0.5$, $\xi =0.4$ and $g_{-2}(\pi )=-0.88$. Here, the curves are shown for $r\geq 2$. But the inner contours in these figures continue to the region with $r<2$ which is not shown and so, are connected vertically artificially. The reason we exclude an inner part goes to back to the validity of our expansions for the radial distances greater than the accretion radius.

Like LS,  there is no restriction on $g_{-2}(\pi )$ in the magnetized case as long as we do not extend the solutions to the inner part of flow or else adopting a model for the accretion and fixing the accretion rate. LS followed the second approach which is much easier and avoids complexities related to the conditions on the sonic surface. Their accretion model is the classical Bondi-Hoyle \citep*{bondi,BondiHoyle}
 model for which there is an analytical relation for accretion rate as,
\begin{equation}\label{eq:Bondi-Hoyle}
\dot{M}= \frac{2 (\lambda^2 + \beta^2)^{1/2}}{(1+\beta^2 )^2}.
\end{equation}

Having the above the relation for the accretion rate, one boundary condition is given at $\theta =0$ as $f_{0}(0)=\dot{M}$. The rest of boundary conditions are at $\theta = \pi$ as $f_{0}(\pi)=f_{0}'(\pi)=0$. Thus, we can solve the second-order equations as a two-points boundary value problem. As far as we know, for the magnetized accretion there is not a closed analytical formula for the accretion rate. However, it was shown that the magnetic field effects suppress the accretion rate  \citep[e.g.,][]{Igu,shadmehri2004,shch,IguNar,Perna2003}. For example, when the accretion occurs as a radiatively  inefficient flow,  the magnetic field is amplified as the gas flows in, and so, the accretion rate is reduced in comparison to the Bondi rate as studied numerically by \citet{IguNar}. But there are many uncertainties regarding to the magnetized accretion, in particular about the true physical mechanism of the accretion or when the accretor is in traveling. In order to proceed analytically, we introduce a toy accretion rate as

\begin{figure}
\epsfig{figure=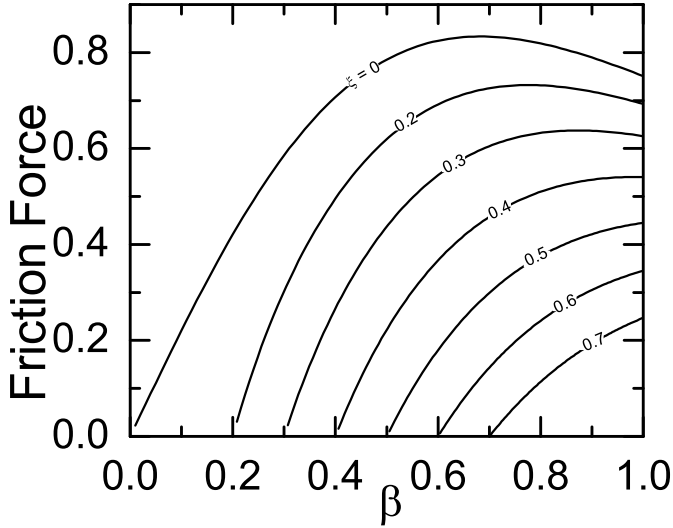,angle=0,scale=0.5}
\caption{The nondimensional dynamical friction force versus  the Mach number $\beta$ for different values of $\xi$. Each curve is limited by the inequality $\xi < \beta$ and the numbers are the corresponding value of $\xi$. }
\label{fig:force}
\end{figure}

\begin{equation}\label{eq:Mdot-mag}
{\dot M}_{\rm mag}= \frac{2 (\lambda^2 + \beta^2 )^{1/2}}{(1+\xi^2 + \beta^2)^2},
\end{equation}
which gives a  reduced rate because of the magnetic field.
 Here, the rate decreases with $\xi$ and it gives Bondi-Hoyle accretion rate if we set $\xi =0$. It is as if we replace the thermal pressure by the sum of thermal and the magnetic pressures, i.e. $c_{\rm s}^2 \rightarrow c_{\rm s}^{2}+v_{\rm A}^{2}$. In fact, the extra pressure provided by the magnetic force lead to the reduction in the accretion rate. We emphasize our prescribed accretion rate relation (\ref{eq:Mdot-mag}) is {\it not} an accurately physically confirmed relation, but it shows a reduction to the accretion rate with increasing the strength  of the magnetic field at least qualitatively.

Having equation (\ref{eq:Mdot-mag}) for the accretion rate, we did a detailed parameter study by solving the main equations (\ref{eq:second-r-2}) and (\ref{eq:second-theta-2}) numerically. We also obtained the non-magnetic results of LS by setting $\xi =0$. In particular, LS proved  that the nondimensional dynamical friction force for a non-magnetized flow is
\begin{equation}
F=-\int_{0}^{\pi} \left [ (1-\beta^2) \sin\theta \cos\theta g_{-2} + \beta (1+ \cos^2 \theta ) f'_{0} \right ] d\theta,
\end{equation}
and then they showed analytically that the above integral is equal to $\beta \dot{M}$. They obtained the above relation by integrating the stress tensor over sphere surrounding the object. Since we are considering magnetic effects, it is known that the contribution of the magnetic field to the stress tensor is $\frac{1}{2}B^2 \delta_{ij}-B_{i}B_{j}$, i.e. the tress tensor is
 \begin{equation}
 T = \rho {\bf U} {\bf U} + (P + \frac{1}{2}B^2) {\bf I} - {\bf B}{\bf B}.
 \end{equation}
Thus, in our non-dimensional notation, the magnetic part of the drag force becomes
\begin{equation}
F_{\rm mag}= - \frac{\xi^2}{2} \int b^2 r^2 \cos\theta \sin\theta d\theta + \xi^2 \int b_{\rm z} b_{\rm r} r^2 \sin\theta d\theta.
\end{equation}
Also, we have
\begin{equation}
b_{\rm z} b_{\rm r} r^2 \sin\theta = \frac{\cot\theta}{r^2} \left ( \frac{\partial\Phi }{\partial\theta }\right )^2 + \frac{1}{r} \left ( \frac{\partial\Phi}{\partial\theta } \right ) \left ( \frac{\partial\Phi }{\partial r} \right ),
\end{equation}
and
\begin{equation}
b^2 r^2 \sin\theta \cos\theta = \cot\theta \left [ \frac{1}{r^2} \left (\frac{\partial\Phi}{\partial\theta } \right )^2 + \left (\frac{\partial\Phi}{\partial r} \right )^2 \right ].
\end{equation}

Now, we can substitute corresponding expansions into the above equations to calculate the drag force. However, the resulting integral is not integrable and one should calculate it numerically. Our numerical solutions show that the dynamical friction force in the magnetized case is approximately equal to $(\beta - \xi^2 / \beta ) f_0 (0)$, where the accretion rate is replaced by the magnetized accretion rate, i.e. equation (\ref{eq:Mdot-mag}). Thus, we numerically confirm $F \simeq (\beta - \xi^2 / \beta ) \dot{M}_{\rm mag}$. Figure \ref{fig:force} shows dynamical friction force $F$ versus parameter $\beta$ for different values of $\xi$. Note that our solutions are restricted to the cases with $\xi < \beta$. Each curve is labeled by its parameter $\xi$ and it is truncated due to the inequality $\xi < \beta$.

\section{conclusions}
When a gravitating object move  subsonically through a magnetized medium, we studied the structure of the flow  using a perturbation analysis. Our study is a direct generalization of LS to the magnetized case. We found analytical relations for the first-order density and velocity perturbations. According to these solutions, the structure of the flow in comparison to the non-magnetized medium is less dense.  However, in order to determine the dynamical friction force the second-order variables are needed. The dynamical friction force was found to be $F \simeq \beta \dot{M}_{\rm mag}$ which implies a reduced force because of considering the magnetic effects. Is this relation valid for a physically confirmed magnetized accretion rate? This is an open question to be studied further. Future models of the magnetized accretion should give us an analytical or numerical relation for the accretion rate as a function of $\beta$ and $\xi$, i.e. $\dot{M}_{\rm mag}(\beta , \xi )$. Then, our main equations (\ref{eq:second-r-2}) and (\ref{eq:second-theta-2}) could be solved easily to calculate the dynamical friction force. However, we think the magnetic effects will still reduce the dynamical friction force.

In the cluster of the galaxies, the motion of the each member of group is affected by the dynamical friction force and it has been suggested that work done by the force may heat up the intragalactic medium \citep*{kim2004,kim2005}. This extra source of the heating may resolve the cooling flow problem in the cluster of the galaxies (for a detailed analysis, see \citet*{kim2005}).  \citet*{kim2005} used the classical formula for calculating the drag force which is based on assumption that the object's radius is greater than the accretion radius. But as LS argued in such systems, accretion radius is comparable to the radius of the accretor. Thus, it would be interesting to calculate the heating due to the friction force using the formula that is presented by LS, at least to the regions of the cluster where the galaxies are moving nearly subsonically like at the outer parts. But intracluster medium is magnetized, and so, our analysis imply a lower heating rate when magnetic fields are taken into account. We finally note that several issues of potential importance were  not investigated  in this paper. For example, we assumed that the ambient medium is isothermal. One can relax this assumption and re-do a similar analysis, but for a magnetized isentropic ambient medium.

\section*{Acknowledgments}

We are grateful to the referee, Steven Stahler,   for his helpful comments and suggestions that improved the paper.

\bibliographystyle{mn2e}
\bibliography{DFmag}

\end{document}